\documentclass[a4paper,11pt]{article}
\pdfoutput=1 

\usepackage{jheppub} 

\usepackage[T1]{fontenc} 
\usepackage{graphicx}

\usepackage{MnSymbol}
\usepackage{empheq}
\usepackage{slashed} 
\usepackage{amssymb}
\usepackage{amsmath}

\newcommand{\DD}{\Delta}

\usepackage{tikz}
\usetikzlibrary{decorations}

\title{\boldmath Relativistic partial waves for celestial amplitudes}

\author{Y. T. Albert Law,$^1$}
\author{Michael Zlotnikov,$^1$}

\affiliation{$^1$ Department of Physics, Center for Theoretical Physics,
Columbia University, 538 West 120th Street, New York, NY 10027, USA.}
\emailAdd{yal2109@columbia.edu}
\emailAdd{mz2737@columbia.edu}

\abstract{\\
The formalism of relativistic partial wave expansion is developed for four-point celestial amplitudes of massless external particles. In particular, relativistic partial waves are found as eigenfunctions to the product representation of celestial Poincar\'e Casimir operators with appropriate eigenvalues. The requirement of hermiticity of Casimir operators is used to fix the corresponding integral inner product, and orthogonality of the obtained relativistic partial waves is verified explicitly. The completeness relation, as well as the relativistic partial wave expansion follow. Example celestial amplitudes of scalars, gluons, gravitons and open superstring gluons are expanded on the basis of relativistic partial waves for demonstration. A connection with the formulation of relativistic partial waves in the bulk of Minkowski space is made in appendices.
}

\begin{document} 
\maketitle
\flushbottom

\section{Introduction}

Celestial amplitudes describe quantum field theoretic scattering on the celestial sphere at light-like infinity as opposed to the familiar scattering amplitudes in Minkowski space. The focus on the two-dimensional celestial sphere as the underlying space in this description of four-dimensional scattering leads to a notion of holography \cite{deBoer:2003vf}. It is of interest to determine so called Celestial Conformal Field Theories (CCFT) living on the celestial sphere, which would give rise to celestial amplitudes without reference to their Minkowski space counterparts. Furthermore, this boundary parametrization is particularly convenient, since it equally well can be used to describe scattering in asymptotically flat space-times \cite{Bondi:1962px,Sachs:1962wk}.

The map to the celestial sphere for Minkowski space scattering amplitudes with external particles of various masses and (bosonic) spins has been worked out in \cite{Pasterski:2016qvg,Pasterski:2017kqt,Law:2020tsg}, and the map was performed for particular amplitude examples in \cite{Pasterski:2016qvg,Lam:2017ofc,Banerjee:2017jeg,Nandan:2019jas,Pasterski:2017ylz,Schreiber:2017jsr,Stieberger:2018edy,Puhm:2019zbl,Guevara:2019ypd,Albayrak:2020saa}.
A somewhat different variation of this map was considered in \cite{Banerjee:2018gce,Banerjee:2018fgd,Banerjee:2019prz}.
Soft theorems familiar from Minkowski space amplitudes correspond to so called conformal soft theorems for celestial amplitudes, which were studied in \cite{Donnay:2018neh,Banerjee:2019aoy,Fan:2019emx,Pate:2019mfs,Nandan:2019jas,Adamo:2019ipt,Puhm:2019zbl,Guevara:2019ypd,Himwich:2019dug}. 
The representation of BMS symmetry generators on the celestial sphere, and other aspects such as OPE expansions of celestial operators were discussed in \cite{Ball:2019atb,Fotopoulos:2019tpe,Pate:2019lpp,Banerjee:2020kaa,Fotopoulos:2019vac,Fotopoulos:2020bqj,Fan:2020xjj}. Information contained at past boundary of future null infinity concerning local excitations in an asymptotically flat bulk space-time has been investigated in \cite{Laddha:2020kvp}. Double copy construction of celestial amplitudes was outlined in \cite{Casali:2020vuy}. In \cite{Law:2019glh,Law:2020tsg} the authors of the current work tabulated the explicit constraints on celestial amplitudes implied by Poincar\'e symmetry (based on conformal Ward identities and momentum conservation), making use of the representation of Lorentz generators and massless \cite{Stieberger:2018onx}, as well as newly established massive momentum generators on the celestial sphere. 

In attempts to better understand the particle or operator content of the holographic theory giving rise to celestial amplitudes, conformal OPE limits mentioned above, or conformal partial wave decomposition of celestial amplitudes \cite{Lam:2017ofc,Nandan:2019jas} have been considered. These techniques are relevant to the framework of representation theory of the 2D global conformal group (homogeneous Lorentz group in the bulk) or its local extension, which sufficiently parametrizes scattering in a particular Lorentz frame without allowing translations out of that frame. Just as in Minkowski space, restricting to such a Lorentz frame may conveniently simplify the calculations in a scattering problem.

However, since our aim is to better understand the underlying theory, effectively gauge fixing part of the symmetry of the problem prior to setting up orthogonal modes may not help our cause. Minkowski space translations are part of the bulk isometry group, and therefore part of the symmetry group of a holographic dual -- and thus must leave their footprint in the unitary representation theory chosen to parametrize the problem at hand. This suggests spanning orthogonal modes in the holographic dual in terms of unitary Poincar\'e representation theory, as we will discuss in more detail in a later section. Therefore, as we proposed in \cite{Law:2019glh}, instead of conformal partial wave decomposition we consider relativistic partial wave decomposition, the derivation of which on the celestial sphere is the subject of this work. Relativistic partial waves in the bulk of Minkowski space have been considered a long time ago, see \cite{Joos:1962qq,Macfarlane:1962zza} or \cite{Moussa:1966gjd} for an early review. In this paper we re-derive the corresponding results in the celestial formulation of quantum scattering with four massless external particles.

In section \ref{sec:CPWspin} we recall the parametrization of momenta in the bulk, the map of amplitudes to the celestial sphere, as well as celestial representations of Poincar\'e symmetry generators.

In section \ref{sec:repTheory} we emphasize the importance of using unitary Poincar\'e representation theory on the celestial sphere, or appropriate BMS representation theory when working with asymptotically flat space-times in the bulk  \cite{Bondi:1962px,Sachs:1962wk,Barnich:2009se,McCarthy:1972wk,McCarthy:1973qvg}.

In section \ref{sec:RPW} we recall Poincar\'e Casimir operators, their product representation, and derive relativistic partial waves as Casimir operator eigenfunctions with appropriate eigenvalues. We fix the required integral inner product by the hermiticity condition of Casimir operators, and demonstrate orthogonality of relativistic partial waves under this inner product. Finally, we establish a completeness relation and relativistic partial wave expansion.

In section \ref{sec:examples} we demonstrate the explicit relativistic partial wave expansion of celestial scalar, gluon, graviton, and open superstring gluon amplitudes.

In appendix \ref{app:bulkRPW} we connect the celestial relativistic partial waves to their bulk representation, and in appendix \ref{app:bulkRPWA3A3} we outline how relativistic partial waves are related to a combination of two three-point amplitudes.

\section{Preliminaries}
\label{sec:CPWspin}
\subsection{Kinematic definitions and amplitude map to the celestial sphere}
Mass $m$ on-shell particle momentum $p^2=-m^2$ in Minkowski space can be parametrized as
\begin{align}
\label{mp}
p^\mu = m \hat p^{\mu}(y,z,\bar{z})\equiv m\left(\frac{1+y^2+z\bar{z}}{2y},\frac{\bar{z}+z}{2y},i\frac{\bar{z}-z}{2y},\frac{1-y^2-z\bar{z}}{2y}\right),
\end{align}
where $y,z,\bar z$ variables live on Euclidean AdS, whose $SL(2,\mathbb{C})$ isometry corresponds to Lorentz transformations acting on $p^\mu$
\begin{align}
z\to \frac{(a z+b)(\bar{c}\bar{z}+\bar{d})+a\bar{c} y^2}{(c z+d)(\bar{c} \bar{z}+\bar{d})+c \bar{c} y^2}~,~~\bar{z}\to \frac{(\bar{a} \bar{z}+\bar{b})(c  z+  d)+\bar{a}  c y^2}{(\bar{c}\bar{z}+\bar{d})( c z+d)+c \bar{c}y^2}~,~~y\to \frac{y}{(c z+d)(\bar{c}\bar{z}+\bar{d})+c\bar{c} y^2},
\end{align}
with $\bar a=a^*,\bar b=b^*,\bar c=c^*,\bar d=d^*,\in\mathbb{C}$ and $ad-bc=\bar a\bar d-\bar b\bar c=1$. 
A massless, light-like $p^2=0$ momentum in Minkowski space can be written as
\begin{align}
\label{wq}
p^\mu=\omega q^{\mu}(w,\bar w)&\equiv\omega (1+w \bar{w},~\bar{w} + w,~i(\bar{w}-w),~1-w \bar{w}),
\end{align}
with an energy scale $\omega$, while $w,\bar w$ describe a point on the celestial sphere. 
 The same $SL(2,\mathbb{C})$ Lorentz transformations act as M\"obius transformations on $w,\,\bar w$ 
\begin{align}
\label{mobius}
w\to\frac{a w+b}{c w+d}~~,~~~\bar{w}\to\frac{\bar{a} \bar{w}+\bar{b}}{\bar{c} \bar{w}+\bar{d}}~~~,~~~\omega\to|cw+d|^2\omega\,.
\end{align}
Amplitudes are mapped to the celestial sphere by the Pasterski-Shao-Strominger (PSS) prescription \cite{Pasterski:2016qvg,Pasterski:2017kqt} with extension to massive spinning external legs in \cite{Law:2020tsg}
\begin{align}
\label{massAmap}
{A_n}_{J_1,...,J_j}=\left(\prod_{l}\int_0^\infty d\omega_l\,\omega_l^{i\lambda_l}\right)\left(\prod_{i=1}^j \int \frac{dy_i}{y^3_i}dz_i d\bar z_i \sum_{b_i=-s_i}^{s_i}G^{(s_i)}_{J_i b_i}\right){\mathcal{A}_n}_{b_1,...,b_j},
\end{align}
for massless (product over $l$) and massive (product over $i$) external particles, with spin $s_i$ integration weight matrices $G^{(s_i)}_{J_ib_i}$ for massive particles.\footnote{We do not specify matrices $G^{(s_i)}_{J_ib_i}$ here, since they are not going to be required in what follows. The interested reader is invited to see the details in \cite{Law:2020tsg}.} Here, $\lambda_i\in\mathbb{R}$ are parameters in the continuous series representation of conformal operator dimensions $\Delta_j=1+i\lambda_j$, and $J_j$ are operator spins. The resulting object $A_n$ is called a celestial amplitude, and transforms as a correlator of conformal primaries under $SL(2,\mathbb{C})$ transformations on the celestial sphere.

\subsection{Poincar\'e algebra and generator representations on the celestial sphere}
\label{sec:generators}
Massless momentum operator representation on the celestial sphere is given by \cite{Stieberger:2018onx}
\begin{align}
\label{Pml}
P^{\mu}=q^\mu e^{\partial_\Delta}.
\end{align}
Lorentz generators $M^{\mu\nu}=-M^{\nu\mu}$ on the celestial sphere have also been determined in \cite{Stieberger:2018onx}
\begin{align}
M^{01}&=\frac{i}{2}  \left(\left(\bar{w}^2-1\right)
   \partial_{\bar w}+\left(w^2-1\right)
   \partial_{w}+(\Delta-J) \bar{w}+(\Delta+J) w
   \right)\,,\\
	M^{02}&=\frac{-1}{2}  \left(\left(\bar{w}^2+1\right)
   \partial_{\bar w}-\left(w^2+1\right)
   \partial_{w}+(\Delta-J) \bar{w}-(\Delta+J) w
   \right)\,,\\
	M^{03}&=i \left(\bar{w}  \partial_{\bar w}+w
   \partial_w+\Delta
   \right) ~~~~~~,~~~~~~
	M^{12}=\left(-\bar{w}  \partial_{\bar w}+w
   \partial_w+J\right)
   \,,\\
	M^{13}&=\frac{i}{2}  \left(\left(\bar{w}^2+1\right)
   \partial_{\bar w}+\left(w^2+1\right)
   \partial_{w}+(\Delta-J) \bar{w}+(\Delta+J) w
   \right)\,,\\
	M^{23}&=\frac{-1}{2}  \left(\left(\bar{w}^2-1\right)
   \partial_{\bar w}-\left(w^2-1\right)
   \partial_{w}+(\Delta-J) \bar{w}-(\Delta+J) w
   \right)\,.
\end{align}
Together, these operators properly close the Poincar\'e algebra
\begin{align}
\left[M^{\mu\nu},M^{\rho\sigma}\right]&=-i\left(\eta^{\mu\sigma}M^{\nu\rho}+\eta^{\nu\rho}M^{\mu\sigma}-\eta^{\mu\rho}M^{\nu\sigma}-\eta^{\nu\sigma}M^{\mu\rho}\right),\\
\left[M^{\mu\nu},P^{\rho}\right]&=i\left(\eta^{\mu\rho}P^{\nu}-\eta^{\nu\rho}P^{\mu}\right)~~~,~~~
\left[P^{\mu},P^{\nu}\right]=0\,.
\end{align}

\section{Choice of representation theory: $\Delta,J$ vs. $m,s$}
\label{sec:repTheory}
In the following sections we aim to establish an ortho-normal basis of functions, that will allow us to decompose any relativistic four-point celestial amplitude of massless external particles in terms of a sequence or continuum of auxiliary single particle exchanged modes. Therefore, we must pick a representation theory in order to label the exchanged auxiliary modes by appropriate unique characteristic numbers.

Since celestial amplitudes are set up such that they transform as correlators of conformal primaries under Lorentz transformations, one is led to consider the 2D (global) conformal group representation theory on the celestial sphere, which in the case at hand corresponds to labeling exchanged auxiliary modes by the characteristic numbers
\begin{align}
\text{conformal dimension: }\Delta\,,\text{ and spin: }J\,.
\end{align}
Considering the Lorentz group in isolation,\footnote{By, e.g., fixing a Lorentz frame and considering the scattering process without allowing translations out of this frame.} a notion of orthogonality can certainly be established for modes of unique $\Delta,J$ values.

However, a valid holographic description on the celestial sphere must accommodate the same amount of symmetry as the bulk isometry group, namely the Poincar\'e group, and therefore translations must be included. Minkowski space translation generators (momentum operators (\ref{Pml})) are not part of the 2D conformal group generators, and induce real valued shifts in conformal dimensions $\Delta$.  Such shifts move $\Delta$ away from the vertical line $\Delta=1+i\lambda$ with $\lambda\in\mathbb{R}$, and thus destroy the hermiticity of the quadratic conformal Casimir operator (which requires the Casimir eigenvalue $C_{\Delta,J}=\Delta(2-\Delta)-J^2$ to be real). Additionally, conformal partial wave orthogonality is formulated for dimensions in continuous series representation $\Delta=1+i\lambda$ where $\lambda\in\mathbb{R}$, importantly with Re$(\Delta)=1$ fixed \cite{Lam:2017ofc,Nandan:2019jas}. This issue can tentatively be resolved by recalling that any $\Delta\in\mathbb{C}$ dimension values can be expressed in terms of continua of continuous series representation dimensions $\Delta=1+i\lambda$, as was discussed in \cite{Donnay:2020guq} (which however leads to non-vanishing overlap between two modes of different $\Delta$). However, we expect a translation to merely induce a coordinate shift in a mode of the theory, but not to transform one mode into a continuum of different modes. The above discussion boils down to the fact that ($\Delta,J$) does not label unitary irreducible representations of the full Poincar\'e group, and is not sufficient to fully characterize a relativistic orthogonal mode.

As was worked out by Wigner a long time ago \cite{Wigner:1939cj} (see also references within), unitary irreducible representations of the Poincar\'e group are labeled by characteristic numbers
\begin{align}
\text{mass: }m\,,\text{ and spin: }s\,.
\end{align}
In the following we will use these characteristic numbers to label orthogonal exchanged modes.\footnote{External particles of course are also labeled by their mass $m$ and spin $s$ (helicity $J$ in case of $m=0$). Note that conformal primary wavefunctions on the celestial sphere in fact parametrize modes of different $m,s$ (or $m=0$ and $J$) characteristic numbers, since they are properly invertible back to their plane wave counterparts in the bulk of Minkowski space and thus carry the same information.} From this point of view, $\Delta$ along with spin $J$ and the coordinates $z,\bar{z}$ on the celestial sphere, should be thought of as kinematic variables analogous to the usual on-shell four momenta $p^\mu$ and the spin component in direction of propagation.\footnote{This is not surprising, seeing as the celestial rewriting of the Klein-Gordon inner product \cite{Pasterski:2017kqt,Donnay:2020guq} treats conformal dimensions and celestial sphere coordinates on the same footing as a celestial counterpart of momentum variables. In the massless case we can think of the Mellin transform map to the celestial sphere $\int_0^\infty d\omega \omega^{\Delta-1}$ as exchanging the energy scale variable $\omega$ for a dual energy variable $\Delta$. The alternative map to the celestial sphere introduced in \cite{Banerjee:2018fgd} transforms a massless external leg as $\int_0^\infty dE E^{\Delta-1}e^{i E u}$ with $E=(1+w\bar w)\omega$, such that the energy variable $E$ is exchanged for a pair of dual energy variables $\Delta$ and $u$. The two maps are related (up to rescaling) by taking the limit $u\to 0$. Whether we choose to use one dual energy variable or a pair, from the point of view of Poincar\'e group representation theory the resulting conformal primary wavefunction parametrizes a mode of characteristic numbers $m=0$ and helicity $J$.} Real valued shifts of $\Delta$ create no tensions with Poincar\'e representation theory.

The upshot of this discussion is that whenever we attempt to attribute special operator dimension meaning to conformal dimensions on the celestial sphere, we need to keep in mind that we are implicitly truncating the symmetry group of the theory, and therefore are not describing the full quantum field theory in the bulk. Including all relativistic generators, the representation theory changes, and different values of celestial conformal dimensions cease to have particular meaning while each individual orthogonal mode of the theory can take on all conformal dimension values, similarly as it can take on any celestial sphere coordinate values.

Even though the Lorentz group is a subgroup of the Poincar\'e group, the representation theories of these two groups are clearly very different. Considering scattering at light-like infinity of an asymptotically flat space-time instead of Minkowski space, the Poincar\'e group is superseded by the (extended) BMS group \cite{Bondi:1962px,Sachs:1962wk,Barnich:2009se}. Then, in analogy to the above discussion, orthogonal modes better be set up with respect to (extended) BMS representation theory \cite{McCarthy:1972wk,McCarthy:1973qvg} instead of, e.g., the representation theory of the conformal subgroup. This makes the study of irreducible representations of 4D \textit{extended} BMS group in Minkowski signature a very important task, being a prerequisite for understanding orthogonal modes in theories with extended BMS symmetry.

\section{Relativistic partial waves}
\label{sec:RPW}
As emphasized in the previous section, in relativistic quantum theory particles are described by irreducible representations of the Poincar\'e group \cite{Wigner:1939cj}. Each irreducible representation is labeled by two characteristic numbers: mass and spin. 

We may consider a process in which two particles interact and produce a third particle. Group theoretically, the possible types of particles that may be produced in this interaction are governed by the expansion of the direct product of two irreducible representations in terms of a sum of single irreducible representations for all kinematically admissible resulting mass and spin values, with appropriate weights. Similarly, each of the possibly produced auxiliary particles may then decay into two new particles. Mathematically, this similarly takes a weighted sum of single irreducible representations into a new direct product of two irreducible representations of definite masses and spins.

Since all kinematically admissible exchanged auxiliary particles can be exhaustively labeled by different characteristic numbers mass and spin, this suggests the existence of a basis of functions describing each exchange mode separately with appropriate orthogonality and completeness properties. Such functions are called partial waves. In the following we develop the formalism of relativistic partial waves on the celestial sphere for four-point scattering processes with massless external particles.

\subsection{Casimir operators and characteristic numbers}
The characteristic numbers mass and spin of a particle can be retrieved by applying so called Casimir operators that commute with all Poincar\'e algebra generators and possess appropriate eigenvalues. 

The quadratic Poincar\'e Casimir operator is momentum squared, with the obvious eigenvalue negative mass squared
\begin{align}
\label{PP}
P^\mu P_\mu=-m^2.
\end{align}
In case when the particle of interest is massive, the Pauli-Lubanski pseudo-vector
\begin{align}
W^\mu=\frac{1}{2}\epsilon^{\mu\nu\rho\sigma}M_{\nu\rho}P_\sigma,
\end{align}
with Levi-Civita tensor density $\epsilon^{\mu\nu\rho\sigma}$, squares to the quartic Poincar\'e Casimir operator with eigenvalue
\begin{align}
\label{WW}
W^\mu W_\mu = m^2 s(s+1),
\end{align}
where $s$ is the spin characteristic number. In case of a massless particle, $W^\mu$ becomes proportional to $W^\mu\propto P^\mu\propto q^\mu$. Considering that $(-1/2\partial_w\partial_{\bar w}q^\mu)q_\mu=1$, we can extract the helicity characteristic number as
\begin{align}
\label{vW}
\left(-\frac{1}{2}\partial_w\partial_{\bar w}q^\mu\right) W_\mu = J.
\end{align}
Note that when we are dealing with a direct product of two particle representations, momentum operators and Lorentz operators act in a sum, so that
\begin{align}
\label{PP12}
P_{ij}^\mu = P_i^\mu +P_j^\mu~~~\Rightarrow~~~P_{ij}^\mu P_{ij}{}_\mu=-m^2,
\end{align}
as well as
\begin{align}
\label{WW12}
\begin{split}
&M_{ij}{}^{\mu\nu}=M_{i}{}^{\mu\nu}+M_{j}{}^{\mu\nu}\\
&W_{ij}^\mu =\frac{1}{2}\epsilon^{\mu\nu\rho\sigma}M_{ij}{}_{\nu\rho}P_{ij}{}_\sigma
\end{split}~~~\Rightarrow~~~W_{ij}^\mu W_{ij}{}_\mu=m^2 s(s+1),
\end{align}
where $m$ and $s$ here are mass and spin characteristic numbers of the particle produced in the interaction of particles $i$ and $j$. It is easy to check that this readily holds true for the interaction of two spinning massless particles producing a spinning massive particle that was discussed in \cite{Law:2020tsg}.

\subsection{Relativistic partial waves as solutions to Casimir differential equations}
For single particle representations, the Casimir equations (\ref{PP}),(\ref{WW}),(\ref{vW}) automatically evaluate to the respective eigenvalues if we employ the momentum and Lorentz generators defined in section \ref{sec:generators}. However, in the product representation case, the left hand sides of equations (\ref{PP12}) and (\ref{WW12}) become non-trivial operators. Four-point structures that are eigenfunctions of these operators with appropriate Casimir eigenvalues are precisely the relativistic partial waves we seek to obtain. 

Additionally, relativistic partial waves must satisfy all Poincar\'e symmetry requirements, so that without loss of generality we employ the most general Poincar\'e invariant massless external leg four-point structure determined in \cite{Law:2019glh} as an ansatz for the relativistic partial wave
\begin{align}
\label{A4ans}
A_4&=R_{\Delta_i,J_i}\delta\left(i\bar z - i z\right)\tilde f_{\Delta_1,\Delta_2,\Delta_3,\Delta_4}^{J_1,J_2,J_3,J_4}(z ,\bar z)\\
R_{\Delta_i,J_i}&=\frac{\left(\frac{w_{14}}{w_{13}}\right)^{h_3-h_4}
   \left(\frac{w_{24}}{w_{14}}\right)^{h_1-h_2}}{w_{12}^{h_1+h_2} w_{34}^{h_3+h_4}} 
   \frac{\left(\frac{\bar{w}_{14}}{\bar{w}_{13}}\right)^{\bar{h}_{3}-\bar{h}_4}
   \left(\frac{\bar{w}_{24}}{\bar{w}_{14}}\right)^{\bar{h}_1-\bar{h}_2}}{\bar{w}_{12}^{\bar{h}_1+\bar{h}_2}\bar{w}_{34}^{\bar{h}_3+\bar{h}_4}} (z-1)^{\frac{h_1-h_2-h_3+h_4}{2}}(\bar z-1)^{\frac{\bar h_1-\bar h_2-\bar h_3+\bar h_4}{2}},\notag
\end{align}
where $\tilde f$ for now is an arbitrary function of conformal cross ratios $z,\bar z$ defined in (\ref{zzbar}), and $\tilde f$ may have arbitrary dependence on the combination of conformal dimensions $\sum_i\Delta_i$, or dependence on individual $\Delta_i$ that is periodic under $\Delta_i\to\Delta_i+1$. Above, we make use of abbreviations $w_{ij}=w_i-w_j$ and $\bar w_{ij}=\bar w_i-\bar w_j$, conformal cross ratios
\begin{align}
\label{zzbar}
z=\frac{w_{12}w_{34}}{w_{13}w_{24}}~~~,~~~\bar z=\frac{\bar w_{12}\bar w_{34}}{\bar w_{13}\bar w_{24}}\,,
\end{align}
 as well as conformal weights
\begin{align}
h_i=\frac{\Delta_i+J_i}{2}~~~,~~~\bar h_i=\frac{\Delta_i-J_i}{2}\,,
\end{align}
with conformal dimension $\Delta_i=1+i\lambda_i$ (and $\lambda_i\in\mathbb{R}$) and spin $J_i$ on the celestial sphere.

Recall that, as first observed in \cite{Pasterski:2017ylz}, thanks to overall momentum conservation delta functions
\begin{align}
\delta^{(4)}(\sum_i\epsilon_i\omega_i q_i^\mu)~~~ \text{with particle $i$ incoming/outgoing:}~~~\epsilon_i=\pm 1,
\end{align}
the map of any four-point amplitude with massless external legs to the celestial sphere can only have support on the following regions in the conformal cross ratio $z$, depending on whether incoming and outgoing external particles form $s$-, $t$-, or $u$-channel configurations
\begin{align}
\label{schan}
\epsilon_1=\epsilon_2=-\epsilon_3=-\epsilon_4~~~:~~~12\leftrightarrow 34~~~&\Rightarrow~~~1<z,\\
\label{tchan}
\epsilon_1=\epsilon_3=-\epsilon_2=-\epsilon_4~~~:~~~13\leftrightarrow 24~~~&\Rightarrow~~~0<z<1,\\
\label{uchan}
\epsilon_1=\epsilon_4=-\epsilon_2=-\epsilon_3~~~:~~~14\leftrightarrow 23~~~&\Rightarrow~~~z<0.
\end{align}
The expression (\ref{A4ans}) is consistent with all three channels. It is our task to determine the appropriate $\tilde f$ function corresponding to the relativistic partial wave in each of the $z$-regions.

\paragraph{Momentum squared Casimir equation}$~$\\
Acting on the four-point structure (\ref{A4ans}), equation (\ref{PP12}) leads to the requirement
\begin{align}
P_{ij}^\mu P_{ij}{}_\mu A_4=-m^2 A_4\,,
\end{align}
with exchanged mode of mass $m$,\footnote{Since we will be integrating over all mass modes $0<m<\infty$ in all three channels, here we treat $m$ as a dummy variable without introducing different $m$ values in the three channels.} and indices $(ij)\in\{(12),(13),(14)\}$ depending on the channel. This implies the following shift equation for $\tilde f$ in (\ref{A4ans})
\begin{align}
C_2\tilde f_{\Delta_1,\Delta_2,\Delta_3,\Delta_4}^{J_1,J_2,J_3,J_4}(z ,\bar z)=-m^2 \tilde f_{\Delta_1,\Delta_2,\Delta_3,\Delta_4}^{J_1,J_2,J_3,J_4}(z ,\bar z),
\end{align}
with the effective quadratic Casimir operator in the appropriate regions
\begin{align}
\label{C2def}
C_2\equiv\left\{\begin{matrix}
-4e^{\partial_{\Delta_1}+\partial_{\Delta_2}} &;& 1<z \\
-\frac{4}{\sqrt{z \bar z}} e^{\partial_{\Delta_1}+\partial_{\Delta_3}} &;& 0<z<1 \\
-\frac{4\sqrt{(1-z)(1-\bar z)}}{\sqrt{z \bar z}} e^{\partial_{\Delta_1}+\partial_{\Delta_4}}&;& z<0
\end{matrix}\right..
\end{align}
Rescaling $\tilde f$ as follows without loss of generality\footnote{Here, non-periodic dependence on $\Delta_i$ is introduced in combination $\sum_{j=1}^4\Delta_j$, as required by Poincar\'e symmetry \cite{Law:2019glh}.}
\begin{align}
\label{mSolve}
\tilde f_{\Delta_1,\Delta_2,\Delta_3,\Delta_4}^{J_1,J_2,J_3,J_4}(z ,\bar z)=\left(\frac{m}{2}\right)^{-4+\sum_{j=1}^4\Delta_j}\left\{
\begin{matrix}
f_{\vec J,s,\vec\Delta}^{(12)}(z ,\bar z) &;& 1<z \\
(z\bar z)^{\frac{-4+\sum_i\Delta_i}{4}}f_{\vec J,s,\vec\Delta}^{(13)}(z ,\bar z) &;& 0<z<1 \\
\left(\frac{z\bar z}{(1-z)(1-\bar z)}\right){}^{\frac{-4+\sum_i\Delta_i}{4}}f_{\vec J,s,\vec\Delta}^{(14)}(z ,\bar z)&;& z<0
\end{matrix}
\right.,
\end{align}
where $\vec J$ denotes the collection of spins of external particles $(J_1,J_2,J_3,J_4)$, and $\vec\Delta$ analogously, we find that the resulting Casimir equations for $f_{\vec J,s,\vec\Delta}^{(ij)}(z ,\bar z)$ simplify and imply periodicity of $f_{\vec J,s,\vec\Delta}^{(ij)}(z ,\bar z)$ under simultaneous shift $\Delta_i\to\Delta_i+1$ and $\Delta_j\to\Delta_j+1$ with appropriate indices $(ij)\in\{(12),(13),(14)\}$. 

We expect relativistic partial waves to be invertible back to Minkowski space, which implies that we must choose the periodic dependence on each $\Delta_a$ to be trivial (constant in all $\Delta_a$), to make sure that the Mellin-like inverse transform integral converges, as discussed at the end of section 4 in \cite{Law:2019glh}. Therefore, $f^{(ij)}_{\vec J,s}(z ,\bar z)$ carries no $\vec\Delta$ labels from now on.

Since the two incoming massless particles necessarily produce a massive exchanged mode, the second Casimir operator we consider next is the square of Pauli-Lubanski pseudo-vector.

\paragraph{Pauli-Lubanski squared Casimir equation}$~$\\
Acting on the four-point structure (\ref{A4ans}), equation (\ref{WW12}) leads to the requirement
\begin{align}
W_{ij}^\mu W_{ij}{}_\mu A_4=m^2 s(s+1) A_4,
\end{align}
with exchanged mode of mass $m$ and spin $s$, and indices $(ij)\in\{(12),(13),(14)\}$ for appropriate regions in $z$.\footnote{As with mass parameter $m$, here we use spin $s$ as the same dummy variable in all three channels. The channel distinction will enter through different summation ranges over $s$.} The operators $W_{ij}^\mu W_{ij}{}_\mu $ contain derivatives that act on $\delta\left(i\bar z - i z\right)$ within $A_4$. Assuming the parametrization
\begin{align}
\label{ztoxy}
z=x+i\frac{y}{2}~~~,~~~\bar z=x-i\frac{y}{2}~~~,~~~\text{with}~~~x,y\in\mathbb{R}\,,
\end{align}
we note that $\delta'(y)$ and $\delta''(y)$ only have support at $y=0$, just as $\delta(y)$. Expanding their coefficients to leading order around $y=0$, we resolve delta function derivatives in a distributional sense 
\begin{align}
y\delta'(y)=-\delta(y)~~~\text{and}~~~\frac{y^2}{2}\delta''(y)=\delta(y).
\end{align}
With this, the effective Casimir differential equation for $f^{(ij)}_{\vec J,s}(z,\bar z)\to f^{(ij)}_{\vec J,s}(x)$ is given by
\begin{align}
C_4 f_{\vec J,s}(x)=s(s+1)f_{\vec J,s}(x),
\end{align}
with $f_{\vec J,s}(x)=f^{(ij)}_{\vec J,s}(x)$ as introduced in (\ref{mSolve}) involving appropriate indices $i,j$ for the different regions in $x$, and the effective quartic Casimir differential operator
\begin{align}
\label{C4def}
C_4\equiv \left\{
\begin{matrix}
\frac{ x
   \left(\left(\frac{1}{4}\left(J_{12}+J_{34}\right)^2-4\right) x-
   J_{12}J_{34}+10\right)-6}{
   x-1}+(3 x-4)x \partial_x-(x-1) x^2 \partial_x^2&;&1<x\\
	\frac{-\frac{1}{4} \left(J_{13}+J_{24}\right)^2+\left(J_{13}
   J_{24}-6\right) x+2
   x^2+4}{(x-1) x}+(3-2 x) \partial_x +(x-1) x \partial_x^2&;&0<x<1\\
	\frac{(1-x)  \left(\frac{1}{4}
   J_{14}^2 (1-x)+\frac{1}{2} J_{23} J_{14}
   (x+1)+\left(\frac{1}{4}J_{23}^2-4\right) (1-x)\right)}{-x}-3 (x-1)^2 \partial_x+ (x-1)^2 x\partial_x^2&;&x<0
\end{matrix}
\right.,
\end{align}
where we abbreviate $J_{ab}=J_a-J_b$. In all three channels we therefore have hypergeometric differential equations. Independent solutions are as follows, for the $s$-channel with $1<x$
\begin{align}
\label{C4sol12}
&\begin{split}
f^{(12)}_{\vec J,s}(x)=&c_1^{(12)} (x-1)^{\frac{J_{12}-J_{34}}{2}} x^{2-J_{12}} \,
   _2F_1\left({{-s+J_{12},s+J_{12}+1}\atop{J_{12}+J_{34}+1}};\frac{1}{x}\right)\\
	&+c_2^{(12)} (x-1)^{\frac{J_{12}-J_{34}}{2} } x^{s+3} \,
   _2F_1\left({{s+J_{12}+1,s-J_{34}+1}\atop{2 s+2}};x\right)
\end{split},
\end{align}
and for the $t$-channel with $0<x<1$
\begin{align}
\label{C4sol13}
&\begin{split}
f^{(13)}_{\vec J,s}(x)=&c_1^{(13)}  (1-x)^{\frac{J_{13}-J_{24}}{2}
   }x^{2+\frac{J_{13}+J_{24}}{2}} \,
   _2F_1\left({{-s+J_{13},s+J_{13}+1}\atop{J_{13}+J_{24}+1}};x\right)\\
	&+c_2^{(13)}
    (1-x)^{\frac{J_{13}-J_{24}}{2} } x^{1-s+\frac{J_{24}-J_{13}}{2}
   } \,
   _2F_1\left({{s+J_{13}+1,s-J_{24}+1}\atop{2 (s+1)}};\frac{1}{x}\right)
\end{split},
\end{align}
 as well as for the $u$-channel with $x<0$
\begin{align}
\label{C4sol14}
&\begin{split}
f^{(14)}_{\vec J,s}(x)=&c_1^{(14)} 
   \left(1-x\right)^{-J_{23}} (-x)^{2+\frac{J_{14}+J_{23}}{2} }\,
   _2F_1\left({{-s+J_{23},s+J_{23}+1}\atop{-J_{14}+J_{23}+1}};\frac{1}{1-x}\right
   )\\
	&+c_2^{(14)} 
    (1-x)^{s+1}(-x)^{2+\frac{J_{14}+J_{23}}{2} } \,
   _2F_1\left({{s+J_{14}+1,s+J_{23}+1}\atop{2 (s+1)}};1-x\right)
\end{split}.
\end{align}
 The free coefficients $c_1^{(ij)},c_2^{(ij)}$ remain to be set.

Since $s$ is the spin value of a mass $m$ mode exchanged between two incoming massless particles and two outgoing massless particles, the parameter $s$ must take on at least a minimum value to ensure that appropriate couplings through generalized Clebsch-Gordan coefficients group theoretically exist. Therefore, the same restrictions must hold with regard to $s$ and $J_i,J_j$, or $s$ and $\overline{J_i,J_j}$ (outgoing complement), as in the two-massless one-massive three-point structure case discussed in \cite{Law:2020tsg}:
\begin{align}
\label{sreq}
s\geq \max(|J_{ab}|)\equiv\left\{
\begin{matrix}
\max(|J_1-J_2|, |J_3-J_4|)&;&1<x\\
\max(|J_1-J_3|, |J_2-J_4|)&;&0<x<1\\
\max(|J_1-J_4|, |J_2-J_3|)&;&x<0\\
\end{matrix}
\right..
\end{align}
This makes the first parameter in the first hypergeometric function of each solution (\ref{C4sol12}), (\ref{C4sol13}) and (\ref{C4sol14}) a non-positive integer. Therefore, these hypergeometric functions simplify to Jacobi polynomials $P_n^{(a,b)}(z)$ up to some overall constants
\begin{align}
\label{JacobiP}
\, _2F_1(-n,a;b;z)=\frac{\Gamma (b) \Gamma (n+1) }{\Gamma (b+n)}P_n^{(b-1,a-b-n)}(1-2
   z)\,,~~~\text{with}~~~n\in\mathbb{Z}^*.
\end{align}
Note that the second hypergeometric function in each solution (\ref{C4sol12}), (\ref{C4sol13}) and (\ref{C4sol14}) generically evaluates to terms containing logarithms depending on $x$. As we will observe in section \ref{sec:PhiOrtho}, solutions involving such logarithmic terms are not consistently orthogonal for different $s$ values under the required inner product which we will establish in section \ref{sec:InnerProduct}. Additionally, boundary terms arising from the hermiticity condition of quartic Casimir operator (\ref{C4def}) will be required to vanish, which does not consistently hold for contributions (\ref{C4sol12}), (\ref{C4sol13}) and (\ref{C4sol14}) in the inner product of two relativistic partial waves. Therefore, to obtain an orthogonal basis of functions that respects hermiticity of Casimir operators we must restrict the solution space to the Jacobi polynomial part.

With the above, the relativistic partial wave satisfying all symmetry requirements and Casimir differential equations is summarized as (\ref{A4ans}) with (\ref{ztoxy}) and
\begin{align}
\label{RPWs}
\Phi^{\vec J,\DD}_{m,s}(x)&=\tilde f_{\Delta_1,\Delta_2,\Delta_3,\Delta_4}^{J_1,J_2,J_3,J_4}(x)\\
&=\left\{
\begin{matrix}
\sqrt{\frac{(2 s+1) \left(s-J_{12}\right)!
   \left(J_{12}+s\right)!}{m \left(s-J_{34}\right)!
   \left(J_{34}+s\right)!}}\left(\frac{m}{2}\right)^{\DD-4}\frac{ x^{2-J_{12}}}{(x-1)^{\frac{J_{34}-J_{12}}{2} }}
   P_{s-J_{12}}^{\left(J_{12}+J_{34},J_{12}-J_{34}\right)}\left(1-\frac{2}{x}\right)&;& 1<x \\
\sqrt{\frac{(2 s+1)
   \left(s-J_{13}\right)! \left(J_{13}+s\right)!}{m
   \left(s-J_{24}\right)! \left(J_{24}+s\right)!}}\left(\frac{m}{2}\right)^{\DD-4} \frac{x^{\frac{\Delta +J_{13}+J_{24}}{2}
   } }{(1-x)^{\frac{J_{24}-J_{13}}{2} }}
   P_{s-J_{13}}^{\left(J_{13}+J_{24},J_{13}-J_{24}\right)}(1-2 x)&;& 0<x<1 \\
	\sqrt{\frac{(2 s+1) \left(s-J_{23}\right)!
   \left(J_{23}+s\right)!}{m \left(s-J_{14}\right)!
   \left(J_{14}+s\right)!}}\left(\frac{m}{2}\right)^{\DD-4} \frac{(-x)^{\frac{\Delta
   +J_{14}+J_{23}}{2} } }{(1-x)^{\frac{\Delta }{2}+J_{23}-2}}
   P_{s-J_{23}}^{\left(-J_{14}+J_{23},J_{14}+J_{23}\right)}\left(\frac
   {x+1}{x-1}\right)&;& x<0 
\end{matrix}\right.
\,,\notag
\end{align}
where the particular normalizations chosen will ensure appropriate normalization of the orthogonality relation discussed in section \ref{sec:PhiOrtho}. As before, we abbreviate $J_{ij}=J_i-J_j$, the collection of the four incoming and outgoing particle helicity values $ J_1,J_2,J_3,J_4$ is denoted as $\vec J$, and $\DD=\sum_{j=1}^4\Delta_j$.

\subsection{Inner product}
\label{sec:InnerProduct}
In order to establish a notion of completeness and orthogonality of the relativistic partial waves, we must define an appropriate inner product. The function $\tilde f$ is the only factor in (\ref{A4ans}) that contains non-trivial information about the scattering process, so that we choose to set up the inner product such that it acts directly in the space of relativistically non-trivial functions $\tilde f$. We expect the inner product to integrate out all kinematic quantities that are not characteristic numbers, therefore we take the ansatz:
\begin{align}
\label{innerPr}
\langle F,G \rangle_{ij} \equiv \int_{4-i\infty}^{4+i\infty}  \frac{d\DD}{2\pi i} \int_{\mathcal{R}_{ij}} dx \mu_{\vec J,\DD}(x) F \bar G\,,~~~\text{with regions}~~~\left\{\begin{matrix}
\mathcal{R}_{12}&:&1<x<\infty\\
\mathcal{R}_{13}&:&0<x<1\\
\mathcal{R}_{14}&:&-\infty<x<0
\end{matrix}\right.,
\end{align}
for some functions $F$ and $G$, where $\bar G$ is the complex conjugate of $G$, once again $\DD=\sum_{j=1}^4\Delta_j$, and $\mu_{\vec J,\DD}(x)$ is an integration weight that must be fixed such that the inner product respects hermiticity of the two Casimir operators $C_2$ and $C_4$ in $(\ref{C2def})$ and $(\ref{C4def})$
\begin{align}
\label{IPhermiticity}
\langle C_a F,G \rangle_{ij} = \langle  F, C_a G \rangle_{ij}~~~\text{for }a=2,4.
\end{align}
Note that the two kinematic variables $\DD$ and $x$ are enough to trace over the entire kinematic space of the scattering. This is similar to using just two independent Mandelstam variables in Minkowski space to exhaustively parametrize the four-point scattering kinematics.

For $C_2$ in $(\ref{C2def})$, the condition (\ref{IPhermiticity}) implies the requirement
\begin{align}
\mu_{\vec J,\DD+2}(x)=\mu_{\vec J,\DD}(x),~~~\text{for all}~~~-\infty<x<\infty\,,
\end{align}
obtained from Taylor-expanding the (complex conjugated) effective shift operator $e^{-2\partial_{\DD}}$ acting on $\bar G$, and partially integrating in all $\DD$ derivatives.\footnote{Here and in what follows we usually require that functions $F$ and $G$ are such that the integrals converge, and all boundary contributions to the integrals resulting from partial integration cancel out or vanish.} We look for the simplest solution to this equation, so that we set $\mu$ to be independent of $\DD$.

For $C_4$ in $(\ref{C4def})$, in the region $1<x$ the condition (\ref{IPhermiticity}) implies the requirements
\begin{align}
\label{mureq1}
0&=x \left((8-9 x) \mu '(x)-(x-1) x \mu ''(x)\right)+(6-12 x) \mu(x),\\
\label{mureq2}
0&=-2 (x-1) x \left(x \mu '(x)+6 \mu (x)\right),\\
\label{mureq3}
0&=\left[(x-1) x^2 \mu (x)\left( G(x)  F'(x)- F(x)  G'(x)\right)\right]|^\infty_1\,,
\end{align}
while in the region $0<x<1$ the requirements are
\begin{align}
\label{mureqt1}
0&=-(x-1) x \mu ''(x)+(5-6 x) \mu '(x)-4 \mu (x),\\
\label{mureqt2}
0&=-2 (x-1) \left(x \mu '(x)+4 \mu (x)\right),\\
\label{mureqt3}
0&=\left[(x-1)x \mu (x)\left( F(x)  G'(x)- G(x)  F'(x)\right)\right]|^1_0\,,
\end{align}
and in the region $x<0$ the requirements are
\begin{align}
\label{murequ1}
0&=(1-x) \left((x-1) x \mu ''(x)+(9 x-5) \mu '(x)\right)+2 (5-6 x) \mu (x),\\
\label{murequ2}
0&=-2 (x-1) \left((x-1) x \mu '(x)+(6 x-4) \mu (x)\right),\\
\label{murequ3}
0&=\left[(x-1)^2 x \mu (x)\left(F(x)  G'(x)-G(x)  F'(x)\right)\right]|^0_{-\infty}\,.
\end{align}
The first two equations in each case are the coefficients of $F\bar G$ and $F'\bar G$ in the respective $x$-regions, obtained after (doubly) partially integrating in $x$ on the right hand side of (\ref{IPhermiticity}) and canceling terms from the left hand side. Since $F\bar G$ and $F'\bar G$ generically are linearly independent, both coefficients must vanish separately. Here, we dropped potential $\vec J$ labels on $\mu(x)$, since all resulting differential equations  are independent of them. The first order differential equations (\ref{mureq2}), (\ref{mureqt2}) and (\ref{murequ2}) each have a unique solution
\begin{align}
\label{muWeight}
\mu(x)=\left\{
\begin{matrix}
\frac{1}{x^6}&;&1<x\\
\frac{1}{x^4}&;&0<x<1\\
\frac{1}{(1-x)^2x^4}&;&x<0
\end{matrix}
\right.,
\end{align}
which automatically satisfies the second order differential equations (\ref{mureq1}), (\ref{mureqt1}) and (\ref{murequ1}), and is therefore consistent with them. Thus, the integration weight $\mu$ in the inner product definition (\ref{innerPr}) is given by (\ref{muWeight}).

The equations (\ref{mureq3}), (\ref{mureqt3}) and (\ref{murequ3}) impose boundary vanishing conditions on $F$ and $G$ from partial integrations. Note that for an inner product of two relativistic partial waves (\ref{RPWs}) these boundary vanishing requirements are properly satisfied. Had we included contributions (\ref{C4sol12}), (\ref{C4sol13}) and (\ref{C4sol14}) in the partial wave definition, then these boundary vanishing requirements would not be consistently satisfied for all spin values, and would break hermiticity of the quartic Casimir operator. Therefore, logarithmic hypergeometric contributions are properly excluded from the definition of the relativistic partial wave (\ref{RPWs}).

\subsection{Orthogonality and completeness of relativistic partial waves}
\label{sec:PhiOrtho}
With the appropriate inner product (\ref{innerPr}) established, we use it to demonstrate orthogonality and completeness of relativistic partial waves (\ref{RPWs}) with different exchanged characteristic numbers $m_1,s_1$ and $m_2,s_2$, which we denote $\langle \Phi^{\vec J,\DD}_{m_1,s_1},\Phi^{\vec J,\DD}_{m_2,s_2} \rangle$.

The integral over $\DD$ is essentially an inverse Mellin transform of a constant, which readily evaluates to
\begin{align}
\int_{4-i\infty}^{4+i\infty}  \frac{d\DD}{2\pi i} \left(\frac{m_1}{2}\right)^{-4+\DD}\left(\frac{m_2}{2}\right)^{4-\DD}=m_1\delta(m_1-m_2).
\end{align}
Note that due to the simple complex conjugation prescription $\Delta \to 8-\Delta$ in one of the partial waves, power cancellations cause the $x$ and $\Delta$ dependence to factorize for the inner product integral in all three channels.

To perform the $x$ integral, we take the change of variables 
\begin{align}
\begin{matrix}
1-\frac{2}{x}=t,&dx=\frac{2dt}{(t-1)^2}&\text{ so that }&1<x&\text{ maps to }&-1<t<+1,\\
2x-1=t,&dx=\frac{dt}{2}&\text{ so that }&0<x<1&\text{ maps to }&-1<t<+1,\\
\frac{x+1}{1-x}=t,&dx=\frac{2dt}{(t+1)^2}&\text{ so that }&x<0&\text{ maps to }&-1<t<+1.
\end{matrix}
\end{align}
With this, the integrals in each region, up to some constant factors, reduce to the orthogonality relation for Jacobi polynomials (with appropriate integer parameters $a,b,l,n$)\footnote{Various special function identities or integrals used throughout this work can be found, e.g., in \cite{Gradshteyn:2007}.}
\begin{align}
\label{orthoRel}
\int_{-1}^1 (1-t)^a (t+1)^b P_l^{(a,b)}(t) P_n^{(a,b)}(t) \,
   dt=\frac{2^{a+b+1} \Gamma (a+n+1) \Gamma (b+n+1) }{n!
   (a+b+2 n+1) \Gamma (a+b+n+1)}\delta _{l,n},
\end{align}
which holds for integer $n,l$ and with real parts of $a,b>-1$. In our case, parameters $a,b$ possibly can take on negative integer values, in which case we first have to make use of the appropriate integer parameter identities
\begin{align}
P_n^{(a,b)}(t)&=\frac{ \Gamma (a+n+1) \Gamma (b+n+1)
   P_{a+n}^{(-a,b)}(t)}{2^{-a} (t-1)^{a}\Gamma (n+1) \Gamma (a+b+n+1)}\,,~P_n^{(a,b)}(t)=\frac{ \Gamma (a+n+1) \Gamma (b+n+1)
   P_{b+n}^{(a,-b)}(t)}{2^{-b} (t+1)^{b}\Gamma (n+1) \Gamma (a+b+n+1)}\,,\notag\\
P_n^{(a,b)}(t)&=\frac{\Gamma (a+n+1) \Gamma (-a-b-n)
   P_{-a-b-n-1}^{(a,b)}(t)}{\Gamma (n+1) \Gamma (-b-n)}\,,~P_n^{(a,b)}(t)=2^{a+b} (t-1)^{-a} (t+1)^{-b} P_{a+b+n}^{(-a,-b)}(t)\,,
\end{align}
to make upper parameters in both polynomials $P$ non-negative, before applying the orthogonality relation (\ref{orthoRel}). Note that the gamma functions may diverge at non-positive integer values individually, but their ratios are always finite.

With our choice of normalization in (\ref{RPWs}), all gamma functions (factorials) and other factors cancel exactly and each channel produces a Dirac and Kronecker delta function, so that we conclude for the inner product of two relativistic partial waves
\begin{align}
\label{YYortho}
\langle \Phi^{\vec J,\DD}_{m_1,s_1},\Phi^{\vec J,\DD}_{m_2,s_2} \rangle_{ij} = \delta(m_1-m_2)\delta_{s_1,s_2},
\end{align}
which confirms appropriate orthogonality in the exchanged characteristic numbers for each channel $(ij)\in\{(12),(13),(14)\}$ independently.

It is easy to verify for some explicit examples (e.g. $s_1=3$ and $s_2=5$) that had we involved the second hypergeometric functions in the solutions (\ref{C4sol12}), (\ref{C4sol13}) and (\ref{C4sol14}) in the definition of the partial wave as well, then the resulting functions generically would not be orthogonal under the required inner product (\ref{innerPr}).

Recalling (\ref{sreq}), the completeness relation is then given by
\begin{align}
\label{YYcoplete}
\sum_{s\geq\max(|J_{ab}|)}\int_0^\infty dm\frac{\mu(x')}{2\pi i }\Phi^{\vec J,\DD}_{m,s}(x){\overline{\Phi^{\vec J,\DD'}_{m,s}(x')}} = \delta(x-x')\delta(\Delta-\Delta'),
\end{align}
which can be verified by multiplying (\ref{YYcoplete}) with $\Phi^{\vec J,\DD'}_{m',s'}(x')$, integrating over $x'\in\mathbb{R}$ as well as $\DD'\in(4-i\infty,4+i\infty)$, and using (\ref{YYortho}) to recover $\Phi^{\vec J,\DD}_{m',s'}(x)$ on both sides of the equation.

\subsection{Relativistic partial wave expansion}
With the relativistic partial waves (\ref{RPWs}), $\max(|J_{ab}|)$ defined in (\ref{sreq}), inner product (\ref{innerPr}), (\ref{muWeight}), and completeness relation (\ref{YYcoplete}) in hand, it is straightforward to write down the relativistic partial wave expansion
\begin{align}
\label{RPWexp}
\tilde f_{\vec J,\DD}(x)=\sum_{s\geq\max(|J_{ab}|)}\int_0^\infty dm \,\Phi^{\vec J,\DD}_{m,s}(x)\,\langle \tilde f_{\vec J,\DD'}(x'),\Phi^{\vec J,\DD'}_{m,s}(x') \rangle_{ij}\,,
\end{align}
with $\vec J=(J_1,J_2,J_3,J_4)$ and $\DD=\sum_{j=1}^4\Delta_j$, for any function $\tilde f_{\vec J,\DD}(x)$ in the functional space of appropriate symmetry.
The inner product $\langle \tilde f_{\vec J,\DD'}(x'),\Phi^{\vec J,\DD'}_{m,s}(x') \rangle_{ij}$ is chosen appropriately to the region of interest in $x$, and determines the spectrum of masses $m$ and spins $s$ group-theoretically exchanged in the scattering process.

\section{Relativistic partial wave expansion examples}
\label{sec:examples}
We emphasize, that since any relativistic four-point celestial amplitude with massless external legs can be written in the form (\ref{A4ans}), the relativistic partial wave expansion formalism is set up to directly act in the space of relativistically non-trivial functions $\tilde f$ multiplying the terms $R_{\Delta_i,J_i}\delta\left(i\bar z - i z\right)$ which are guaranteed by symmetry. 

In this section we demonstrate explicit relativistic partial wave decompositions of some known example celestial amplitudes. Since the three different channels of the expansion always are to be treated analogously, for brevity we present the calculations only in one channel.

In the explicit examples we consider below, we make use of the Jacobi polynomial integral identity \cite{Gradshteyn:2007}
\begin{align}
\label{PintID}
\int_1^\infty dx\,\frac{ (x-1)^{\sigma }}{x^{\rho +\sigma +2}}  P_n^{(a,b)}\left(\frac{x-2}{x}\right)=\frac{\Gamma (\rho +1) \Gamma (\sigma +1) \Gamma (a+n+1) }{n! \Gamma (a+1) \Gamma (\rho +\sigma +2)}\, _3F_2\left({{-n,a+b+n+1,\rho +1}\atop{a+1,\rho +\sigma +2}};1\right),
\end{align}
which holds for $\sigma>-1$ and $\rho>-1$.

\subsection{Scalar celestial amplitude}
A tree-level amplitude $A_4^{scalar}$ of $(1,2)$ incoming and $(3,4)$ outgoing massless scalars, exchanging a massive scalar of mass $m_s$ via $s$-, $t$- and $u$-channel diagrams was mapped to the celestial sphere in \cite{Nandan:2019jas}. Its celestial relativistically non-trivial part in the sense of (\ref{A4ans}), up to an overall constant, is given by (where we denote $\DD=\sum_{j=1}^4\Delta_j$ as usual)
\begin{align}
\label{fscalar}
\tilde f_{scalar}(x)=&\frac{\pi}{8}\frac{\left(\frac{m_s}{2}\right)^{\Delta -6}}{\sin \left(\frac{\pi  \Delta
   }{2}\right)} x^2  \left(e^{\frac{i \pi  \Delta }{2}}+x^{\frac{\Delta
   }{2}-2}+\left(\frac{x}{x-1}\right)^{\frac{\Delta
   }{2}-2}\right) ~~~\text{for}~~~1<x\,,\\
	=&\frac{1}{m_s^2}\int_0^\infty d\omega\,\omega ^{\Delta -5} \left((x-1)^2\right)^{\frac{\Delta -4}{4}}
	\left(\frac{x^2 }{1-\frac{4 (x-1) \omega ^2}{m_s^2}}
	+ \frac{x^2 }{1+\frac{4 (x-1) \omega ^2}{x m_s^2}}
	+\frac{x^2 }{1+\frac{4 (x-1)^2 \omega ^2}{x m_s^2}}\right)\,,\notag
\end{align}
where in the second line we reverted the result back to a PSS prescription Mellin-transform shape, which is more convenient to perform the $\Delta$ integration in the inner product.
We denote the contributions from the three summands as $\tilde f_{scalar,i}$ with $i=1,2,3$.

This result is expanded in terms of relativistic partial waves (\ref{RPWexp}) in the $s$-channel as
\begin{align}
\label{fscalarExp}
\tilde f_{scalar}(x)=\sum_{s\geq 0}\int_0^\infty dm\,\Phi^{\vec J,\DD}_{m,s}(x)\,\langle \tilde f_{scalar},\Phi^{\vec J,\DD'}_{m,s} \rangle_{12}\,~~~\text{for}~~~1<x.
\end{align}
For the calculation of the inner product (\ref{innerPr}), the first integral over $\Delta$ is the same for all three terms in (\ref{fscalar}):
\begin{align}
\label{omegaint}
\int_{4-i\infty}^{4+i\infty}\frac{d\Delta}{2\pi i} \omega ^{\Delta -5} \left(\frac{m}{2}\right)^{4-\Delta }
   \left((x-1)^2\right)^{\frac{\Delta -4}{4}}=\delta\left(\omega-\frac{m}{2 \sqrt{x-1}}\right)~~~\text{for}~~~1<x\,,
\end{align}
which evaluates to the Dirac delta function since we are essentially dealing with an inverse Mellin transform of a constant.
 The integral over $\omega$ is then trivial in each case, such that the $s$-channel diagram contribution to the inner product, making use of (\ref{PintID}) and straightforward simplifications, reduces to
\begin{align}
\langle \tilde f_{scalar,1},\Phi^{\vec J,\DD'}_{m,s} \rangle_{12} =&\frac{\sqrt{2 s+1} }{\sqrt{m} 
   \left(m_s^2-m^2\right)} \int_{1}^\infty dx\,  x^{-2}P_s^{(0,0)}\left(\frac{x-2}{x}\right)\\
	=&\frac{\sin (\pi  s)}{\pi s} \frac{\sqrt{2 s+1} }{  \sqrt{m}  (s+1) \left(m_s^2-m^2\right)}\\
		=& \frac{\delta_{s,0}}{ \sqrt{m}( m_s^2-  m^2)}~~~\text{since}~~~\frac{\sin (\pi  s)}{\pi s}=\left\{
		\begin{matrix}
		1&;&s=0\\
		0&;&s>0
		\end{matrix}
		\right.~~~\text{for}~~~s\in\mathbb{Z}^*,
\end{align}
which demonstrates that due to the Kronecker delta $\delta_{s,0}$ the $s$-channel diagram only contains an exchange of scalar $s=0$ relativistic partial waves. The exchanged mass spectrum contains a pole at the particular mass value of the exchanged scalar in the diagram. 

Similarly, the inner product for the $t$-channel diagram contribution amounts to
\begin{align}
&\langle \tilde f_{scalar,2},\Phi^{\vec J,\DD'}_{m,s} \rangle_{12}\\
\label{f2phi}
=& \frac{\sqrt{2 s+1}}{\sqrt{m}}\int_{1}^\infty dx \frac{ P_s^{(0,0)}\left(\frac{x-2}{x}\right)}{ x \left(m^2+x
   m_s^2\right)}\left(=\frac{\sqrt{2 s+1}}{2\sqrt{m}}\int_{-1}^1dt\frac{P_s^{(0,0)}(t)}{ m_s^2-\frac{m^2(t-1)}{2} }\right)\\
	=& \frac{\sqrt{2 s+1}}{\sqrt{m}} \int_{0^--i\infty}^{0^-+i\infty}\frac{du}{2\pi i}\Gamma (-u) \Gamma (u+1)\frac{m^{2 u}}{m_s^{2 u+2}} \int_{1}^\infty dx x^{-u-2} 
   P_s^{(0,0)}\left(\frac{x-2}{x}\right) \\
	=&\frac{\sqrt{2 s+1}}{\sqrt{m}} \int_{0^--i\infty}^{0^-+i\infty}\frac{du}{2\pi i} \frac{m^{2 u}}{m_s^{2u+2}}\frac{  \Gamma (-u) \Gamma (u+1)^2  }{ \Gamma (u+2)}\,
   _3F_2\left({{-s,s+1,u+1}\atop{1,u+2}};1\right)\\
	=&\frac{\sqrt{2 s+1}}{\sqrt{m}} \int_{0^--i\infty}^{0^-+i\infty}\frac{du}{2\pi i} \frac{m^{2 u}}{m_s^{2u+2}}\frac{ \Gamma (u+1)^2  \Gamma (s-u)}{\Gamma (s+u+2)}\\
	=& \frac{\sqrt{\pi }   \Gamma (s+1) }{\sqrt{m}\sqrt{2 s+1} \Gamma
   \left(s+\frac{1}{2}\right)m_s^2}\left(\frac{m}{2 m_s}\right)^{2s}\,
   _2F_1\left({{s+1,s+1}\atop{2 s+2}};-\frac{m^2}{m_s^2}\right)
\end{align}
where in the intermediate steps we made use of the Mellin-Barnes representation
\begin{align}
\frac{1}{x+y} =\int_{0^--i\infty}^{0^-+i\infty}\frac{du}{2\pi i} \Gamma (-u) \Gamma (u+1) x^u y^{-1-u},
\end{align}
and evaluated the final step via contour deformation, collecting all residue contributions at the pole locations $u=s+n$ with $n\in\mathbb{Z}^*$.

Analogously, the inner product for the $u$-channel diagram contribution leads to
\begin{align}
\langle \tilde f_{scalar,3},\Phi^{\vec J,\DD'}_{m,s} \rangle_{12} =&\frac{\sqrt{2 s+1}}{\sqrt{m}}\int_{1}^\infty dx \frac{ P_s^{(0,0)}\left(\frac{x-2}{x}\right)}{ x \left(m^2 (x-1)+x
   m_s^2\right)}=\frac{\sqrt{2 s+1}}{2\sqrt{m}}\int_{-1}^1dt\frac{P_s^{(0,0)}(-t)}{ m_s^2-\frac{m^2(t-1)}{2} }\notag\\
	=&\frac{\sqrt{\pi } (-1)^s \Gamma (s+1) }{\sqrt{m}\sqrt{2 s+1} \Gamma
   \left(s+\frac{1}{2}\right)m_s^2}\left(\frac{m}{2 m_s}\right)^{2s}\,
   _2F_1\left({{s+1,s+1}\atop{2 s+2}};-\frac{m^2}{m_s^2}\right).
\end{align}
where we used $P^{(0,0)}_s(-t)=(-1)^s P^{(0,0)}_s(t)$ to relate this integral to (\ref{f2phi}). 

Combining all three contributions together, the inner product with an $s$-channel relativistic partial wave reads
\begin{align}
\label{scalarPhi}
\langle \tilde f_{scalar},\Phi^{\vec J,\DD'}_{m,s} \rangle_{12} =\frac{\delta_{s,0}}{\sqrt{m} ( m_s^2-  m^2)}+\frac{(1+(-1)^s)\sqrt{\pi }  \Gamma (s+1) }{\sqrt{m}\sqrt{2 s+1} \Gamma
   \left(s+\frac{1}{2}\right)m_s^2}\left(\frac{m}{2 m_s}\right)^{2s}\,
   _2F_1\left({{s+1,s+1}\atop{2 s+2}};-\frac{m^2}{m_s^2}\right).
\end{align}
Note that due to the $(1+(-1)^s)$ factor only modes of even spin $s$ receive a non-vanishing weight. For integer values of $s\geq0$ the resulting hypergeometric function generically evaluates to terms involving a natural logarithm $\ln\left(1+\frac{m^2}{m_s^2}\right)$. Since $m_s\in\mathbb{R}$, the integration range $0<m<\infty$ in (\ref{fscalarExp}) always keeps the argument of the logarithm away from its branch cut.

\subsection{Gluon celestial amplitude}
The tree-level gluon four-point amplitude $A_{4,glu}^{--++}$ with helicities $J_1=J_2=-1$ and $J_3=J_4=+1$ was mapped to the celestial sphere in \cite{Pasterski:2017ylz}. Its celestial relativistically non-trivial part in the sense of (\ref{A4ans}), up to an overall constant, is given by
\begin{align}
\label{fglu}
\tilde f^{--++}_{glu}(x)=\delta(i(4-\DD))\frac{x^3}{x-1}.
\end{align}
Gluons are massless particles, such that the collinear limit (here $x\to 1$) forces a Feynman propagator in a four-point amplitude on shell and causes the amplitude to diverge. This divergence is usually regulated by an $i\epsilon$ prescription\footnote{Such a regulator slightly breaks the conformal symmetry of the tree-level gluon amplitude and is required to properly make sense of the amplitude}. In fact, for the inner product (\ref{innerPr}) of the gluon amplitude (\ref{fglu}) with a relativistic partial wave (\ref{RPWs}) the partial integration boundary contributions (\ref{mureq3}), (\ref{mureqt3}) and (\ref{murequ3}) to the Casimir operator hermiticity condition vanish properly only if the $i\epsilon$ prescription or an analogous regulator is in place. For convenience, in what follows we use a $\epsilon\to 0^+$ parameter in the exponent that sufficiently regulates the divergence for the purposes of the inner product integral.\footnote{At loop-level we expect similar behavior, as long as the on-shell regime of a massless propagator is probed by the integral.}

In the gluon case, the $s$-channel relativistic partial wave expansion (\ref{RPWexp}) reads
\begin{align}
\label{fgluExp}
\tilde f^{--++}_{glu}(x)=\sum_{s\geq 0}\int_0^\infty dm\,\Phi^{\vec J,\DD}_{m,s}(x)\,\langle \tilde f^{--++}_{glu},\Phi^{\vec J,\DD'}_{m,s} \rangle_{12}~~~\text{for}~~~1<x\,.
\end{align}
The inner product (\ref{innerPr}) amounts to the integral
\begin{align}
\langle \tilde f^{--++}_{glu},\Phi^{\vec J,\DD'}_{m,s} \rangle_{12} =&\frac{\sqrt{2 s+1}}{\sqrt{m}}\int_{4-i\infty}^{4+i\infty}  \frac{d\DD}{2\pi i} \delta(i(4-\DD))\left(\frac{m}{2}\right)^{4-\DD}\int_{1}^\infty dx  \frac{P_s^{(0,0)}\left(1-\frac{2}{x}\right)}{x(x-1)^{1-\epsilon }}\\
		=& \frac{\sqrt{2 s+1}}{2 \pi\sqrt{m}} \Gamma (1-\epsilon ) \Gamma (\epsilon ) \,
   _3F_2\left({{-s,s+1,1-\epsilon }\atop{1,1}};1\right)\\
	\label{gluPreInPr}
	\approx&\frac{1}{ \epsilon }\frac{(-1)^s \sqrt{2 s+1} }{2\pi 
   \sqrt{m}}+\frac{(-1)^{s+1} \sqrt{2 s+1} }{\pi  \sqrt{m}} H_s,
\end{align}
with harmonic number $H_s=\sum_{n=1}^s\frac{1}{n}$.
In the first line we introduced a small parameter $\epsilon\to 0^+$ to regulate the integral over $x$, whose evaluation in the second line then follows from (\ref{PintID}). Finally, we expanded the resulting hypergeometric polynomial around small $\epsilon$. 

It may seem problematic that we obtain infinite contributions $1/\epsilon$ under $\epsilon\to 0^+$, which stem from the collinear boundary of integration $x\to 1$. However, it turns out that these infinities cancel out exactly in the sum over all partial waves. This can be seen by making use of the completeness relation for Legendre polynomials
\begin{align}
\sum _{k=0}^{\infty } (2 k+1) P_k^{(0,0)}(x) P_k^{(0,0)}(y)=2 \delta (x-y).
\end{align}
Plugging (\ref{gluPreInPr}) into the expansion (\ref{fgluExp}), the coefficient of the infinite term $1/\epsilon$ is proportional to
\begin{align}
\label{PP0}
\sum _{s=0}^{\infty } (2 s+1) P_s^{(0,0)}\left(\frac{x-2}{x}\right) (-1)^s= \sum _{s=0}^{\infty } (2 s+1) P_s^{(0,0)}\left(\frac{x-2}{x}\right)P_s^{(0,0)}(-1)= \delta \left(1-\frac{1}{x}\right)=0~\text{for}~1<x.
\end{align}
This tells us that in the case at hand, any inner product result with the $s$-dependence $(-1)^s \sqrt{2 s+1}$ is equivalent to zero when used as expansion weights in the relativistic partial wave decomposition (\ref{fgluExp}). Therefore, the only relevant inner product contribution is given by
\begin{align}
\label{gluInPr}
\langle \tilde f^{--++}_{glu},\Phi^{\vec J,\DD'}_{m,s} \rangle_{12} =\frac{(-1)^{s+1} \sqrt{2 s+1} }{\pi  \sqrt{m}} H_s.
\end{align}

\paragraph{Resummation of the relativistic partial wave expansion:}$~$\\
Since we used a somewhat subtle argument to show that infinite terms do not contribute and the inner product simplifies to (\ref{gluInPr}), here we explicitly verify that (\ref{gluInPr}) is indeed the complete partial wave weight.

In order to resum the partial wave expansion and recover the original function $\tilde f$, we use the following integral representation of the harmonic number
\begin{align}
\label{polygamma}
H_s=\int_0^\infty dt\frac{e^{-t}}{1-e^{-t}}+\int_0^\infty dt\frac{e^{- t}(e^{- t})^s}{e^{-t}-1}\,,
\end{align}
where we have intentionally separated the integral into two summands: The first summand is $s$ independent, such that its contribution vanishes in the sum over partial waves as in (\ref{PP0}). Therefore, only the second summand has a non-trivial contribution to the partial wave expansion.  

Furthermore, we make use of the Legendre polynomial summation identity
\begin{align}
\label{PLid}
\sum_{s=0}^\infty (2 s+1) y^s P_s^{(0,0)}(z) =(2 y\partial_y+1) \sum_{s=0}^\infty  y^s P_s^{(0,0)}(z)=(2 y\partial_y+1) \frac{1}{\sqrt{y^2-2 y z+1}}=\frac{1-y^2}{\left(y^2-2 y z+1\right)^{3/2}}\,,
\end{align}
valid for $-1<z<1$ and $|y|<1$.

With this, the expansion (\ref{fgluExp}) with (\ref{gluInPr}) and (\ref{polygamma}) (second summand only) readily resums to
\begin{align}
\tilde f^{--++}_{glu}(x)=&\int_0^\infty \frac{dm}{m}\left(\frac{m}{2}\right)^{\Delta-4} \sum_{s= 0}^\infty \int_0^\infty dt\frac{ (2 s+1) x^2 \left(-e^{-t}\right)^{s+1}
   }{\pi  \left(e^{-t}-1\right)}P_s^{(0,0)}\left(1-\frac{2}{x}\right)\\
	=& \delta (i (4-\Delta ))\int_0^\infty dt\frac{\sqrt{2} x^{7/2} \cosh \left(\frac{t}{2}\right)}{(x \cosh (t)+x-2)^{3/2}}=\delta(i(4-\DD))\frac{x^3}{x-1}~\text{for}~1<x\,.
\end{align}
In the first step we evaluated the $m$ integral, which amounts to a Mellin transform of a constant resulting in $2\pi\delta(i(4-\Delta))$. We also made use of (\ref{PLid}) to perform the sum over $s$, such that we recover the expected original result (\ref{fglu}) from the remaining integral over $t$.

\subsection{Graviton celestial amplitude}
The tree-level graviton four-point amplitude $A_{4,grav}^{--++}$ with helicities $J_1=J_2=-2$ and $J_3=J_4=+2$ was mapped to the celestial sphere in \cite{Stieberger:2018edy,Puhm:2019zbl}. Its celestial relativistically non-trivial part, up to an overall constant, is given by
\begin{align}
\label{fgrav}
\tilde f^{--++}_{grav}(x)&=x^4  (x-1)^{\frac{\Delta -4}{2}}\int_0^\infty \frac{d\omega}{2\pi}\,\omega ^{\Delta -3}=\frac{x^4}{x-1} \int_0^\infty \frac{d\omega}{2\pi}\,\omega ^{\Delta -3}~~~\text{for}~~~1<x\,,
\end{align}
where for convenience we rescaled the integration variable as $\omega\to\omega/\sqrt{x-1}$. 

As in the gluon case, gravitons are massless particles as well and require an analogous regulator near the collinear region $x\to 1$.\footnote{Similar to gluons, tree-level graviton amplitudes in $4D$ Minkowski space have also been observed to feature hidden conformal invariance \cite{Loebbert:2018xce}, when the symmetry is not broken by an $i\epsilon$ prescription.} 

The respective relativistic partial wave expansion (\ref{RPWexp}) reads
\begin{align}
\label{fgravExp}
\tilde f^{--++}_{grav}(x)=\sum_{s\geq 0}\int_0^\infty dm\,\Phi^{\vec J,\DD}_{m,s}(x)\,\langle \tilde f^{--++}_{grav},\Phi^{\vec J,\DD'}_{m,s} \rangle_{12}~~~\text{for}~~~1<x\,.
\end{align}
The inner product (\ref{innerPr}) evaluates analogously to the gluon case
\begin{align}
&\langle \tilde f^{--++}_{grav},\Phi^{\vec J,\DD'}_{m,s} \rangle_{12} \\
=&\frac{\sqrt{2 s+1} }{\sqrt{m}}\int_{4-i\infty}^{4+i\infty}  \frac{d\DD}{2\pi i} \int_0^\infty \frac{d\omega}{2\pi}\, \omega^{\DD-3}\left(\frac{m}{2}\right)^{4-\DD}\int_{1}^\infty dx  \frac{
   P_s^{(0,0)}\left(1-\frac{2}{x}\right)}{x^{2 \epsilon }(x-1)^{1-\epsilon } }\\
=&\frac{\sqrt{2 s+1} }{\sqrt{m}}\int_0^\infty \frac{d\omega}{2\pi}\,\omega\delta\left(\frac{2\omega}{m}-1\right)\cos \left(\frac{\pi  s}{2}\right)\frac{\Gamma (\epsilon )^2 \sin
   \left(\frac{\pi  (2 \epsilon -s)}{2} \right)  \Gamma
   \left(\frac{s+1}{2}\right)  \Gamma
   \left(\frac{s}{2}-\epsilon +1\right)}{2^{2 \epsilon -1}\pi  \Gamma \left(\frac{s}{2}+1\right)
   \Gamma \left(\frac{s}{2}+\epsilon +\frac{1}{2}\right)}\\
\hateq&\frac{\sqrt{2 s+1} }{2\pi\sqrt{m}}\left(\frac{m}{2 }\right)^2 \left((-1)^s+1\right)\frac{  \Gamma
   \left(\frac{s}{2}+\frac{1}{2}\right) \Gamma (\epsilon ) \Gamma
   \left(\frac{s}{2}-\epsilon +1\right)}{2^{2 \epsilon }\Gamma \left(\frac{s}{2}+1\right) \Gamma
   (1-\epsilon ) \Gamma \left(\frac{s}{2}+\epsilon +\frac{1}{2}\right)}\\
\label{gravPreInPr}
\approx&\frac{1}{\epsilon }\frac{\sqrt{2 s+1} }{2\pi\sqrt{m}}\left(\frac{m}{2 }\right)^2 
\left((-1)^s+1\right) 
-\frac{\sqrt{2 s+1} }{\pi\sqrt{m}}\left(\frac{m}{2 }\right)^2  \left((-1)^s+1\right) H_s.
\end{align}
As in the gluon case, we introduced a small parameter $\epsilon\to 0^+$ to regulate the $x$-integral. Making use of (\ref{PintID}), the integral leads to a hypergeometric function that simplifies to a ratio of gamma functions in this case. We notice that the result constains $\cos \left(\frac{\pi  s}{2}\right)$, which reveals that only even spin $s$ contributions are non-vanishing. Therefore, in the next step we simplify with the assumption that $\frac{s}{2}$ is integer, which we relate to previous terms by the equivalence sign $\hateq$ instead of equality. Finally, we expand the result for small $\epsilon$ and find that divergent terms have the $s$-dependence $\left((-1)^s+1\right)\sqrt{2 s+1}$. As in the gluon case this means that all such terms resum to zero in the relativistic partial wave expansion, where in this case apart from (\ref{PP0}) we also employ the Legendre polynomial orthogonality relation limiting to the opposite boundary
\begin{align}
\label{PP10}
\sum _{s=0}^{\infty } (2 s+1) P_s^{(0,0)}\left(\frac{x-2}{x}\right) = \sum _{s=0}^{\infty } (2 s+1) P_s^{(0,0)}\left(\frac{x-2}{x}\right)P_s^{(0,0)}(1)= \delta \left(-\frac{1}{x}\right)=0~\text{for}~1<x<\infty.
\end{align}
This reveals that the only non-trivial contribution to the inner product, when it is considered as a weight in the relativistic partial wave expansion, stems from the finite terms
\begin{align}
\label{gravInPr}
\langle \tilde f^{--++}_{grav},\Phi^{\vec J,\DD'}_{m,s} \rangle_{12} =-\frac{\sqrt{2 s+1} }{\pi\sqrt{m}}\left(\frac{m}{2 }\right)^2  \left((-1)^s+1\right) H_s,
\end{align}
which are non-zero for even exchanged spin $s$ values.

\paragraph{Resummation of the relativistic partial wave expansion:}$~$\\
The integral over $\omega$ in (\ref{fgrav}) is recovered from (\ref{fgravExp}) simply by plugging in (\ref{RPWs}) and (\ref{gravInPr}), and taking the change of integration variable $m=2\omega$. One part of the contribution to the inner product (\ref{gravInPr}) is equivalent to the gluon case treated in the previous sub-section, so that we directly use the corresponding resummation result. Making use of (\ref{polygamma}) and (\ref{PP10}), the remaining contribution evaluates to
\begin{align}
\frac{\tilde f^{--++}_{grav}(x)}{\int \frac{d\omega}{2\pi}\,\omega^{\Delta-3}}=&\frac{x^3}{x-1}+2 \sum_{s= 0}^\infty \int_0^\infty dt\frac{ (2 s+1) x^2 \left(e^{-t}\right)^{s+1}
   }{  \left(1-e^{-t}\right)}P_s^{(0,0)}\left(1-\frac{2}{x}\right)\\
	=&\frac{x^3}{x-1}+ \int_0^\infty dt\frac{\sqrt{2} x^{7/2} \cosh \left(\frac{t}{2}\right)}{(x \cosh (t)-x+2)^{3/2}}\\
	=&\frac{x^3}{x-1}+x^3=\frac{x^4}{x-1}~\text{for}~1<x\,,
\end{align}
where all steps are analogous to the gluon resummation calculation above.

\subsection{Open superstring gluon celestial amplitude}
The tree-level open superstring gluon four-point amplitude $A_{4,str.glu}^{--++}$ with helicities $J_1=J_2=-1$ and $J_3=J_4=+1$ was mapped to the celestial sphere in \cite{Stieberger:2018edy}. Its celestial relativistically non-trivial part, up to an overall constant, is given by
\begin{align}
\label{fstrglu}
\tilde f^{--++}_{str.glu}(x)&=\frac{2}{\pi} \alpha'  x^3  (x-1)^{\frac{\Delta }{2}-2} \int_0^\infty d\omega\,\omega ^{\Delta -3} B\left(4 (x-1)
   \alpha'  \omega ^2,\frac{4 (x-1)^2 \alpha'  \omega ^2}{x}+1\right),
\end{align}
where $B(a,b)$ is the Euler beta function and $\alpha'$ is the 'universal Regge slope' parameter. The gluon field theory situation is recovered as $\alpha'\to 0$.

The $s$-channel relativistic partial wave expansion (\ref{RPWexp}) reads
\begin{align}
\label{fsgluExp}
\tilde f^{--++}_{str.glu}(x)=\sum_{s\geq 0}\int_0^\infty dm\,\Phi^{\vec J,\DD}_{m,s}(x)\,\langle \tilde f^{--++}_{str.glu},\Phi^{\vec J,\DD'}_{m,s} \rangle_{12}~~~\text{for}~~~1<x\,.
\end{align}
The $\Delta$ integral in the inner product (\ref{innerPr}) evaluates analogously to (\ref{omegaint}), making the subsequent $\omega$ integral trivial, while the $x$ integral leads to
\begin{align}
\langle \tilde f^{--++}_{str.glu},\Phi^{\vec J,\DD'}_{m,s} \rangle_{12} =&
\frac{\sqrt{2 s+1}}{2\pi\sqrt{m}}\alpha'    m^{2 } \int_1^\infty dx \frac{P_s^{(0,0)}\left(\frac{x-2}{x}\right) B\left(m^2 \alpha' ,\frac{(x-1)
   \alpha'  m^2}{x}+1\right)}{(x-1)^{1-\epsilon } x^{1+\epsilon }}\\
	=&\frac{\sqrt{2 s+1}}{2\pi\sqrt{m}}\sum_{n=0}^\infty\frac{ m^{2 n+2} \alpha'{} ^{n+1} \Gamma (n+\epsilon )^2
   B^{(0,n)}\left(\alpha'  m^2,1\right)}{\Gamma (n+1) \Gamma (n-s+\epsilon )
   \Gamma (n+s+\epsilon +1)}\\
	\approx&\frac{1}{\epsilon }\frac{(-1)^s \sqrt{2 s+1} }{2\pi\sqrt{m}}+\frac{ (-1)^{s+1} \sqrt{2 s+1} }{\pi\sqrt{m}}H_s\\
	&+\sum_{n=s+1}^\infty\frac{\sqrt{2 s+1} m^{2 n+\frac{3}{2}} \alpha'{} ^{n+1} \Gamma (n)
   B^{(0,n)}\left(\alpha'  m^2,1\right)}{2\pi n \Gamma (n-s) \Gamma (n+s+1)}
\end{align}
where we Taylor expanded the beta function for $0<\frac{x-1}{x}<1$ to use the identity (\ref{PintID}), such that $B^{(0,n)}(a,b)$ is the $n$th derivative of the beta function with respect to the second argument. The first two terms after expansion for small $\epsilon\to 0^+$ stem from the $n=0$ Taylor expansion term, and are therefore equivalent to the gluon result discussed above. As before, for the purposes of the relativistic partial wave expansion, any inner product terms with the $s$-dependence $(-1)^s \sqrt{2 s+1}$ resum to zero (\ref{PP0}) and can be dropped.

Alternatively, we can make use of
\begin{align}
B(a,b)=\sum_{k=0}^\infty\frac{(-1)^k \Gamma (a)}{(b+k) \Gamma (k+1) \Gamma (a-k)}~~\text{so that}~~B^{(0,n)}(a,b)=\sum_{k=0}^\infty\frac{n! (-1)^{k+n} \Gamma (a)  }{(b+k)^{n+1}\Gamma (k+1) \Gamma
   (a-k)}\,,
\end{align}
to perform the sum over $n$ and write the (finite and non-trivial part of the) result as
\begin{align}
\langle \tilde f^{--++}_{str.glu},\Phi^{\vec J,\DD'}_{m,s} \rangle_{12} =&\frac{ (-1)^{s+1} \sqrt{2 s+1} }{\pi\sqrt{m}}H_s\\
	&+\frac{m^{2 s+\frac{7}{2}} \alpha'{} ^{s+2}s! \Gamma \left(m^2 \alpha' \right)}{\sqrt{\pi }2^{2 s+1}\sqrt{2 s+1} \Gamma \left(s+\frac{1}{2}\right)}\sum_{k=0}^\infty \frac{ (-1)^{k+s+1}  \, _2F_1\left({{s+1,s+1}\atop{2 s+2}};-\frac{m^2 \alpha' }{k+1}\right)
     }{ (k+1)^{s+2} k!
    \Gamma \left(m^2 \alpha' -k\right)}.\notag
\end{align}
The remaining sum over $k$ is non-trivial and does not seem to lead to a simple expression in terms of some special function. Note the striking similarity of the hypergeometric function in the final sum to the scalar inner product result (\ref{scalarPhi}), such that the sum over $k$ can be interpreted as contributions from a tower of masses $m_k^2=\frac{1+k}{\alpha'}$. Of course, one can similarly consider heterotic string gluon amplitudes or stringy graviton amplitudes discussed in \cite{Stieberger:2018edy}, which we omit to avoid clutter of examples.

\acknowledgments
We thank A. Joyce for interesting comments and discussions about the draft. AL and MZ are supported by the US Department of Energy under
contract DE-SC0011941.

\appendix

\section{Bulk Minkowski space representation of relativistic partial waves}
\label{app:bulkRPW}
In this work we have derived the relativistic partial waves directly on the celestial sphere. However, their representation in the bulk of Minkowski space can easily be given as well. Making use of Mandelstam variables
\begin{align}
\mathfrak{s}=(\epsilon_1 p_1+\epsilon_2 p_2)^2~~~,~~~\mathfrak{t}=(\epsilon_1 p_1+\epsilon_3 p_3)^2~~~,~~~\mathfrak{u}=(\epsilon_1 p_1+\epsilon_4 p_4)^2\,,
\end{align}
with $\epsilon_i=\pm 1$ for particle $i$ incoming/outgoing, and momenta $p_i^\mu$ in (\ref{wq}), the bulk representation of the relativistic partial wave (\ref{A4ans}) with $\tilde f$ as in (\ref{RPWs}) reads
\begin{align}
\label{bulkPhi}
\Phi^{\vec J,m,s}_{bulk}&=8 m^2\delta^{(4)}\left(\sum_{i=1}^4\epsilon_i p_i\right)\times\\
&\times\left({\scriptstyle{\sqrt{\frac{(2 s+1) \left(s-J_{1,2}\right)!
   \left(J_{1,2}+s\right)!}{m \left(s-J_{3,4}\right)! \left(J_{3,4}+s\right)!}}}} \delta \left(m^2+\mathfrak{s}\right)
   \left(-\frac{\mathfrak{s}}{\mathfrak{t}}\right)^{-J_{1,2}}
   \left(\frac{\mathfrak{s}+\mathfrak{t}}{\mathfrak{t}}\right)^{\frac{J_{1,2}-J_{3,4}}{2}
   } 
   P_{s-J_{1,2}}^{\left(J_{1,2}+J_{3,4},J_{1,2}-J_{3,4}\right)}\left(\frac{2
   \mathfrak{t}+\mathfrak{s}}{\mathfrak{s}}\right) \right.\notag\\
&+{\scriptstyle\sqrt{\frac{(2 s+1) \left(s-J_{1,3}\right)!
   \left(J_{1,3}+s\right)!}{m \left(s-J_{2,4}\right)! \left(J_{2,4}+s\right)!}}}\delta \left(m^2+\mathfrak{t}\right)
   \left(\frac{\mathfrak{s}}{\mathfrak{t}}\right)^{\frac{J_{1,3}+J_{2,4}}{2}}
   \left(\frac{\mathfrak{s}+\mathfrak{t}}{\mathfrak{t}}\right)^{\frac{J_{1,3}-J_{2,4}}{2}} 
   P_{s-J_{1,3}}^{\left(J_{1,3}+J_{2,4},J_{1,3}-J_{2,4}\right)}\left(\frac{2
   \mathfrak{s}+\mathfrak{t}}{\mathfrak{t}}\right)\notag\\
&+	\left. {\scriptstyle \sqrt{\frac{(2 s+1) \left(s-J_{2,3}\right)! \left(J_{2,3}+s\right)!}{m
   \left(s-J_{1,4}\right)! \left(J_{1,4}+s\right)!}}}\delta \left(m^2+\mathfrak{u}\right)
   \left(-\frac{\mathfrak{s}}{\mathfrak{t}}\right)^{\frac{J_{1,4}+J_{2,3}}{2}}
   \left(-\frac{\mathfrak{s}+\mathfrak{t}}{\mathfrak{t}}\right)^{-J_{2,3}}
   P_{s-J_{2,3}}^{\left(J_{2,3}-J_{1,4},J_{1,4}+J_{2,3}\right)}\left(\frac{\mathfrak{s}-\mathfrak{t}}{\mathfrak{s}+\mathfrak{t}}\right)\right)\,.\notag
\end{align}
Performing the Mellin transform map (\ref{massAmap}) to the celestial sphere for this expression is trivial, since the four $\omega_i$ integrations are localizing four of the five delta functions of momentum conservation $\delta^{(4)}\left(\sum_{i=1}^4\epsilon_i p_i\right)$ and respective channel constraints $\delta \left(m^2+\mathfrak{s}\right)$, $\delta \left(m^2+\mathfrak{t}\right)$ or $\delta \left(m^2+\mathfrak{u}\right)$. With appropriate delta function integration Jacobians, (\ref{A4ans}) with (\ref{RPWs}) are directly recovered.

Equation \ref{bulkPhi} is in line with the results obtained in the bulk formalism \cite{Joos:1962qq,Macfarlane:1962zza,Moussa:1966gjd}.

\section{Relativistic partial waves from two three-point amplitudes}
\label{app:bulkRPWA3A3}
In the bulk, relativistic partial waves can be obtained by gluing together two three-point amplitudes with two massless legs and one massive leg, while tracing over the phase space of massive exchange on-shell momentum and spin multiplet. For the $s$-channel, suppressing a spectator normalization factor, this yields\footnote{Both other channel contributions can be treated completely analogously.}
\begin{align}
\label{PhiAAi}
\Phi^{\vec J,m,s}_{bulk,s-ch.}&=2m^4\sum_{b=-s}^s\int_0^\infty\frac{dy}{y^3}\int dz d\bar z \delta^{(4)}\left(\sum_{i=1}^2\epsilon_i p_i+p\right)A_b(p_1,p_2,p)\delta^{(4)}\left(\sum_{j=3}^4\epsilon_j p_j-p\right)A_b^*(p_3,p_4,-p),
\end{align}
where $p_i^\mu$ with $i=1,2,3,4$ are massless momenta of the external particles parametrized as (\ref{wq}), and $p^\mu$ is the massive (on-shell) momentum (\ref{mp}) of the exchanged mode of spin $s$ and mass $m$.\footnote{For scattering in Minkowski signature $(1,3)$, three point amplitudes of massless external particles vanish on-shell, so that the exchanged auxiliary mode is always massive. Considering scattering, e.g., in $(2,2)$ signature, massless three-point amplitudes do not vanish on-shell, so that in this case, apart from massive exchanged modes, also massless exchanged modes would contribute additionally.} Index $b$ on each of the two amplitudes represents the spin polarization along the axis of propagation of the massive exchanged mode, and we are summing over the complete multiplet. The asterisk on $A_b^*$ denotes complex conjugation.

Integration over $y,z,\bar z$ traces out the on-shell phase space available to the massive exchanged momentum $p^\mu$. All three integrations are saturated by three of the delta functions present in the integrand. Additionally, the massive polarization vector $\epsilon^{\mu_1...\mu_s}_b$ can be extracted from the three-point amplitudes, so that we obtain an expression in terms of uncontracted three-point amplitudes
\begin{align}
\label{PhiAA}
\Phi^{\vec J,m,s}_{bulk,s-ch.}&=8m^2\delta^{(4)}\left(\sum_{i=1}^4\epsilon_i p_i\right)\delta\left(m^2+\mathfrak{s}\right)\left(\sum_{b=-s}^s\epsilon^{\mu_1...\mu_s}_b\epsilon^*{}^{\nu_1...\nu_s}_b\right)A_{\mu_1...\mu_s}A^*_{\nu_1...\nu_s}.
\end{align}
The remaining delta functions already resemble the corresponding terms in (\ref{bulkPhi}). The sum over $b$ is evaluated by employing the massive polarization completeness relation for each corresponding spin $s$. The uncontracted amplitudes $A_{\mu_1...\mu_s}$ and $A^*_{\nu_1...\nu_s}$ depend on helicities of the massless external particles. 

To obtain an analogous construction of relativistic partial waves in terms of three-point structures on the celestial sphere, the corresponding integral has the form
\begin{align}
R_{\Delta_i,J_i}\delta\left(i\bar z - i z\right)\Phi^{\vec J,\DD}_{m,s}(x) 
=&\sum_{J=-s}^s\int_{1-i\infty}^{1+i\infty}\frac{d\Delta}{2\pi i}\int dw d\bar w V^\Delta_J(w,\bar w)A_3{}^{\Delta_1,\Delta_2,\Delta}_{J_1,J_2,J}(A_3{}^{\Delta_3,\Delta_4,\Delta}_{J_3,J_4,J})^*\,,
\end{align}
with an appropriate integration weight $V^\Delta_J(w,\bar w)$, that could potentially be determined from Poincar\'e Ward identities. 

An alternative, simpler way to obtain this construction, is to start with a three-point celestial amplitude with two massless external legs and one massive external leg as discussed in \cite{Law:2020tsg}
\begin{align}
\mathcal{A}^{\Delta_i,\Delta_j,\Delta_l}_{J_i,J_j;s_l,J_l}=&\int_0^\infty d\omega_i \omega_i^{\Delta_i-1}\int_0^\infty d\omega_j \omega_j^{\Delta_j-1}\int_{0}^{\infty}\frac{dy_l}{y_l^3}\int  dz_l d\bar z_l \sum_{b=-s_l}^{s_l}G^{(s_l)}_{J_l,b}A_{b}\,,
\end{align}
where $A_{b}$ is the corresponding bulk Minkowski space amplitude, and perform the inverse map back to Minkowski space on the massive leg, as outlined in section 2.5 of \cite{Law:2020tsg}, resulting in
\begin{align}
{\mathcal{\tilde A}}{}^{\Delta_i,\Delta_j}_{J_i,J_j;s_l,b_l}=&\int_0^\infty d\omega_i \omega_i^{\Delta_i-1}\int_0^\infty d\omega_j \omega_j^{\Delta_j-1}A_{b_l}\,.
\end{align}
 Then, taking two of such amplitudes with only massless external legs mapped to the celestial sphere, the gluing proceeds exactly as in (\ref{PhiAAi}) and (\ref{PhiAA}), while the Mellin transforms of the massless legs to the celestial sphere come along as spectators. 

\paragraph{Simple examples:}
In the following we verify the construction (\ref{PhiAAi}) and (\ref{PhiAA}) on three simplest examples of relativistic partial waves with four external massless scalars, and $s=0,1,2$ exchange mode. In case of massless external scalars, the uncontracted amplitudes above read
\begin{align}
\label{uncontrA}
A_{\mu_1...\mu_s}=(\epsilon_1p_1-\epsilon_2p_2)_{\mu_1}...(\epsilon_1p_1-\epsilon_2p_2)_{\mu_s}~~~,~~~A^*_{\nu_1...\nu_s}=(\epsilon_3p_3-\epsilon_4p_4)_{\nu_1}...(\epsilon_3p_3-\epsilon_4p_4)_{\nu_s},
\end{align}
while any occurrence of the massive exchanged momentum $p^\mu$ in the polarization tensor completeness relations is obviously replaced by $p^\mu=\epsilon_3p_3^\mu+\epsilon_4 p_4^\mu=-\epsilon_1p_1^\mu-\epsilon_2 p_2^\mu$.

\paragraph{Scalar external legs and $s=0$ exchange:}
Three point amplitudes of scalars are constants. Replacing polarization tensors and uncontracted amplitudes in (\ref{PhiAA}) by a constant, we trivially find agreement with (\ref{bulkPhi}) up to a normalization factor, since $P_0^{(0,0)}(x)=1$.

\paragraph{Scalar external legs and $s=1$ exchange:}
In case of spin $s=1$ exchange, the massive polarization vector completeness relation reads
\begin{align}
\sum_{b=-1}^1\epsilon^{\mu}_b\epsilon^*{}^{\nu}_b=\eta^{\mu\nu}-\frac{p^\mu p^\nu}{p^2}.
\end{align}
Combined with (\ref{uncontrA}), we therefore get
\begin{align}
\left(\sum_{b=-1}^1\epsilon^{\mu}_b\epsilon^*{}^{\nu}_b\right)(\epsilon_1p_1-\epsilon_2p_2)_\mu(\epsilon_3p_3-\epsilon_4p_4)_\nu&=(\epsilon_1p_1-\epsilon_2p_2)\cdot(\epsilon_3p_3-\epsilon_4p_4)+0\\
&=\mathfrak{s}+2\mathfrak{t}=\mathfrak{s} P_1^{(0,0)}\left(\frac{\mathfrak{s}+2\mathfrak{t}}{\mathfrak{s}}\right).
\end{align}
We seem to have obtained a stray factor of $\mathfrak{s}$ when using this result in (\ref{PhiAA}) compared to (\ref{bulkPhi}). However, note that thanks to the delta function $\delta\left(m^2+\mathfrak{s}\right)$ in (\ref{PhiAA}) we can replace $\mathfrak{s}$ by $-m^2$, such that the difference is in normalization only.

\paragraph{Scalar external legs and $s=2$ exchange:}
In case of spin $s=2$ exchange, the massive polarization vector completeness relation reads
\begin{align}
\sum_{b=-2}^2\epsilon^{\mu_1\mu_2}_b\epsilon^*{}^{\nu_1\nu_2}_b=\frac{1}{2}P^{\mu_1\nu_1}P^{\mu_2\nu_2}+\frac{1}{2}P^{\mu_2\nu_1}P^{\mu_1\nu_2}-\frac{1}{3}P^{\mu_1\mu_2}P^{\nu_1\nu_2}~~~\text{with}~~~P^{\mu\nu}=\eta^{\mu\nu}-\frac{p^\mu p^\nu}{p^2}.
\end{align}
Combined with (\ref{uncontrA}), we find
\begin{align}
&\left(\sum_{b=-2}^2\epsilon^{\mu_1\mu_2}_b\epsilon^*{}^{\nu_1\nu_2}_b\right)(\epsilon_1p_1-\epsilon_2p_2)_{\mu_1}(\epsilon_1p_1-\epsilon_2p_2)_{\mu_2}(\epsilon_3p_3-\epsilon_4p_4)_{\nu_1}(\epsilon_3p_3-\epsilon_4p_4)_{\nu_2}=\\
&=\left(\mathfrak{s}+2\mathfrak{t}\right)^2-\frac{1}{3}\mathfrak{s}^2=\frac{2}{3}\mathfrak{s}^2 P_2^{(0,0)}\left(\frac{\mathfrak{s}+2\mathfrak{t}}{\mathfrak{s}}\right).
\end{align}
In this case we get a stray factor of $\mathfrak{s}^2$. Making use of the delta function $\delta\left(m^2+\mathfrak{s}\right)$ in (\ref{PhiAA}) to replace $\mathfrak{s}^2$ by $m^4$, we find agreement between (\ref{PhiAA}) and (\ref{bulkPhi}) up to normalization.
%
%


%

\end{document}